\documentclass[acmtog, nonacm=true]{acmart}
\acmSubmissionID{263}

\usepackage{booktabs} 
\usepackage{multirow}
\usepackage[ruled]{algorithm2e} 

\SetAlFnt{\small}
\SetAlCapFnt{\small}
\SetAlCapNameFnt{\small}
\SetAlCapHSkip{0pt}

\acmJournal{TOG}
\acmVolume{40}
\acmNumber{4}
\acmArticle{86}
\acmYear{2021}
\acmMonth{8}

\setcopyright{rightsretained}

\acmDOI{10.1145/3450626.3459786}

\graphicspath{{pic/}}

\begin{document}
\title{TransPose: Real-time 3D Human Translation and Pose Estimation with Six Inertial Sensors}

\author{Xinyu Yi}
\affiliation{
  \institution{BNRist and school of software, Tsinghua University}
  \city{Beijing}
  \country{China}
}
\email{yixy20@mails.tsinghua.edu.cn}

\author{Yuxiao Zhou}
\affiliation{
  \institution{BNRist and school of software, Tsinghua University}
  \city{Beijing}
  \country{China}
}
\email{yuxiao.zhou@outlook.com}

\author{Feng Xu}
\affiliation{
 \institution{BNRist and school of software, Tsinghua University}
 \city{Beijing}
 \country{China}
}
\email{xufeng2003@gmail.com}

\begin{abstract}
Motion capture is facing some new possibilities brought by the inertial sensing technologies which do not suffer from occlusion or wide-range recordings as vision-based solutions do. 
However, as the recorded signals are sparse and quite noisy, online performance and global translation estimation turn out to be two key difficulties. 
In this paper, we present TransPose, a DNN-based approach to perform full motion capture (with both global translations and body poses) from only 6 Inertial Measurement Units (IMUs) at over 90 fps. 
For body pose estimation, we propose a multi-stage network that estimates leaf-to-full joint positions as intermediate results. 
This design makes the pose estimation much easier, and thus achieves both better accuracy and lower computation cost. 
For global translation estimation, we propose a supporting-foot-based method and an RNN-based method to robustly solve for the global translations with a confidence-based fusion technique. 
Quantitative and qualitative comparisons show that our method outperforms the state-of-the-art learning- and optimization-based methods with a large margin in both accuracy and efficiency. 
As a purely inertial sensor-based approach, our method is not limited by environmental settings (e.g., fixed cameras), making the capture free from common difficulties such as wide-range motion space and strong occlusion. 

\end{abstract}

%
%

\ccsdesc[500]{Computing methodologies~Motion capture}
%
%

\keywords{IMU, Pose Estimation, Inverse Kinematics, Real-time, RNN}

\begin{teaserfigure}
  \includegraphics[width=\textwidth]{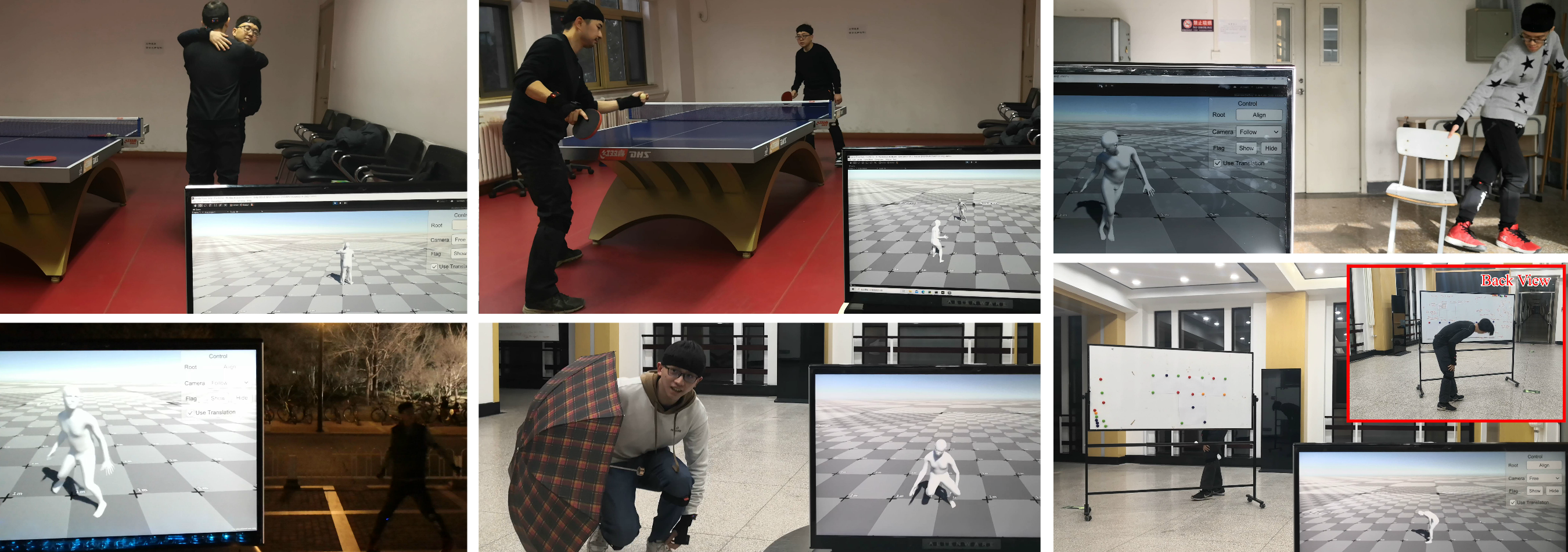}
  \caption{We present a real-time motion capture approach that estimates poses and global translations using only six inertial measurement units. As a purely inertial sensor-based approach, our system does not suffer from occlusion, challenging environment or multi-person ambiguities, and achieves long-range capture with real-time performance.
  }
  \Description{Live demos. Our system handles strong occlusion, dark and outdoor environment, close interaction, and multiple users with high accuracy.}
  \label{fig:teaser}
\end{teaserfigure}

\maketitle

\section{Introduction}
%
Human motion capture (mocap), aiming at reconstructing 3D human body movements, plays an important role in various applications such as gaming, sports, medicine, VR/AR, and movie production.
So far, vision-based mocap solutions take the majority in this topic.
One category requires attaching optical markers on human bodies and leverages multiple cameras to track the markers for motion capture.
The marker-based systems such as Vicon\footnote{\url{https://www.vicon.com/}} are widely applied and considered accurate enough for industrial usages.
However, such approaches require expensive infrastructures and intrusive devices, which make them undesirable for consumer-level usages.
Recently, another category focuses on pose estimation using a few RGB or RGB-D cameras \cite{Trumble2016, Tome2018, Habibie2019, Xiang2019, Mehta2020, Chen2020}.
Although these methods are much more lightweight, they are sensitive to the appearance of humans since distinguishable features need to be extracted from images.
In consequence, these methods usually work poorly for textureless clothes or challenging environment lighting.
Furthermore, all vision-based approaches suffer from occlusion.
Such a problem can sometimes be solved by setting dense cameras (which further makes the system heavy and expensive), but it is often impractical in some applications.
For example, it is very difficult to arrange cameras in a common room full of furniture or objects, which may occlude the subject from any direction.
Another limitation of vision-based approaches is that the performers are usually restricted in a fixed space volume.
For daily activities such as large-range walking and running, carefully controlled moving cameras are required to record enough motion information, which is very hard to achieve \cite{flycap}.
These disadvantages are fatal flaws for many applications and thus lead to limited usability of the vision-based solutions.
%
\par
%
In contrast to vision-based systems, motion capture using body-worn sensors is environment-independent and occlusion-unaware. 
It does not require complicated facility setups and lays no constraint on the range of movements.
Such characteristics make it more suitable for customer-level usages.
As a result, motion capture using inertial sensors is gaining more and more focus in recent years.
Most related works leverage Inertial Measurement Units (IMUs) to record motion inertia, which are economical, lightweight, reliable, and commonly used in a fast-growing number of wearables such as watches, wristbands, and glasses.
The commercial inertial mocap system, Xsens\footnote{\url{https://www.xsens.com/}}, uses 17 IMUs to estimate joint rotations.
Although accurate, the dense placement of IMUs is inconvenient and intrusive, preventing performers from moving freely.
On the other hand, the facility requirements for such a system are still beyond the acceptance of normal customers.
The work of SIP \cite{SIP} demonstrates that it is feasible to reconstruct human motions from only 6 IMUs.
However, as an optimization-based method, it needs to access the entire sequence and takes a long time to process.
The following state-of-the-art work, DIP \cite{DIP}, achieves real-time performance with better quality by leveraging a bidirectional RNN, also using 6 IMUs.
However, it still fails on challenging poses, and the frame rate of 30 fps is not sufficient to capture fast movements, which are very common in practical applications.
More importantly, it only estimates body poses without global movements which are also important in many applications such as VR and AR.
The IMU itself is incapable of measuring distances directly.
Some previous works \cite{Vlasic2007, Liu2011} use additional ultrasonic sensors to measure global translations, which are expensive and subject to occlusion.
Other possible solutions use GPS localization, which is not accurate enough and only works in outdoor capture.
As a result, super-real-time estimation of both body poses and global translations from sparse worn sensors is still an open problem.
%
\par
%
To this end, we introduce our approach, TransPose, which estimates global \textit{translations} and body \textit{poses} from only 6 IMUs at unprecedented 90 fps with state-of-the-art accuracy.
Performing motion capture from sparse IMUs is extremely challenging, as the problem is severely under-constrained.
We propose to formulate the pose estimation task in a multi-stage manner including 3 subtasks, each of which is easier to solve, and then estimate the global translations using the fusion of two different but complementary methods.
As a result, our whole pipeline achieves lower errors but better runtime performance.
Specifically, for \textit{multi-stage pose estimation}, the task of each stage is listed as follows:
\textit{1)} \textit{leaf joint position estimation} estimates the positions of \textit{5 leaf joints} from the IMU measurements;
\textit{2)} \textit{joint position completion} regresses the positions of \textit{all 23 joints} from the leaf joints and the IMU signals;
\textit{3)} \textit{inverse kinematics solver} solves for the joint rotations from positions, i.e. the inverse kinematics (IK) problem.
For \textit{fusion-based global translation estimation}, we combine the following two methods:
\textit{1)} \textit{foot-ground contact estimation} estimates the foot-ground contact probabilities for both feet from the leaf joint positions and the IMU measurements, and calculates the root translations on the assumption that the foot with a larger contact probability is not moving;
\textit{2)} \textit{root velocity regressor} regresses the local velocities of the root in its own coordinate frame from all joint positions and the IMU signals.
%
\par
%
The design of using separate steps in the pose estimation is based on our observation that estimating joint positions as an intermediate representation is easier than regressing the rotations directly.
We attribute this to the nonlinearity of the rotation representation, compared with the positional formulation.
On the other hand, performing IK tasks with neural networks has already been well investigated by previous works \cite{HandIK,holden2018robust,6D,BodyIK}.
However, these works neglect the importance of motion history.
As the IK problem is ill-posed, the motion ambiguity is commonly seen, which can only be eliminated according to motion history.
Here, in each stage, we incorporate a bidirectional recurrent neural network (biRNN) \cite{birnn} to maintain motion history.
This is similar to \cite{DIP}, but due to our 3-stage design, our network is significantly smaller with higher accuracy.
%
\par
%
As for the global translation estimation, the intuition behind is to predict which foot contacts the ground and fix the contacting foot to deduce the root movement from the estimated pose.
Besides, to cope with the cases where no foot is on the ground, we further use an RNN to regress the root translation as a complement.
Finally, our method merges these two estimations to predict the final translation using the fusion weight determined by the foot-ground contact probability.
The inspiration comes from \cite{Zou2020,PhysCap} which leverage foot-ground contact constraints to reduce foot-sliding artifacts in vision-based mocap, but we have a different purpose which is to solve for the translation from the estimated pose.
As shown in the experiments, this method can handle most movements like walking, running, and challenging jumping.
%
\par
%
We evaluate our approach on public datasets, where we outperform the previous works DIP and SIP by a large margin both qualitatively and quantitatively with various metrics, including positional and rotational error, temporal smoothness, runtime, etc.
We also present live demos that capture a variety of challenging motions where the performer can act freely regardless of occlusion or range.
These experiments demonstrate our significant superiority to previous works.
In conclusion, our contributions are:
\begin{itemize}
	\item A very fast and accurate approach for real-time motion capture with global translation estimation using only 6 IMUs.
	\item A novel structure to perform pose estimation which explicitly estimates joint positions as intermediate subtasks, resulting in higher accuracy and less runtime.
	\item The first method to explicitly estimate global translations in real-time from as sparse as 6 IMUs only.
\end{itemize}

\section{Related Work}
%
The research on human motion capture has a long history, especially for the vision-based motion capture. 
Since this is not the category of this work, we refer readers to the surveys \cite{moeslund2006survey,moeslund2001survey} for more information.
In this section, we mainly review the works using inertial sensors (or combined with other sensors) which are closely related to our approach.
%
\subsection{Combining IMUs with Other Sensors or Cameras}
%
Typically, a 9-axis IMU contains 3 components: an accelerometer that measures accelerations, a gyroscope that measures angular velocities, and a magnetometer that measures directions.
Based on these direct measurements, the drift-free IMU orientations can be solved \cite{Foxlin1996, Bachmann2002, Roetenberg2005, Del2018, Vitali2020}  leveraging Kalman filter or complementary filter algorithms.
However, reconstructing human poses from a sparse set of IMUs is an under-constrained problem, as the sensors can only provide orientation and acceleration measurements which are insufficient for accurate pose estimation.
To cope with this difficulty, one category of works \cite{Vlasic2007, Liu2011} propose to utilize additional ultrasonic distance sensors to reduce the drift in joint positions.
In the work of \cite{Liu2011}, database searching is used for similar sensor measurements, and the results help to construct a local linear model to regress the current pose online.
Although the global position can be determined by leveraging the distance sensors, subjects can only act within a fixed volume to keep that the distance can be measured by ultrasonic sensors.
Another category of works propose to combine IMUs with videos \cite{Pons-Moll2010, Pons-Moll2011, Marcard2016, Malleson2017, Malleson2019, Gilbert2018, Henschel2020, Zhang2020}, RGB-D cameras \cite{Helten2013, Zheng2018}, or optical markers \cite{Andrews2016}.
Gilbert et al. \cite{Gilbert2018}  fuse multi-viewpoint videos with IMU signals and estimate human poses using 3D convolutional neural networks and recurrent neural networks.
Global positions can also be determined in some works \cite{Malleson2017, Malleson2019, Andrews2016} due to the additional information from vision.
Although these works achieve great accuracy, due to the need of vision, the capture suffers from occlusion and challenging lighting conditions, and can only be performed within a restricted area to keep the subject inside the camera field of view.
To this end, Marcard et al. \cite{Marcard2018} propose to use a moving camera to break the limitation on the space volume.
Nonetheless, all the methods that require visual or distance sensor inputs are substantially limited by the capture environment and occlusion.
%
\subsection{Methods Based on Pure Inertial Sensors}
%
As opposed to methods based on the fusion of IMUs and other sensors, approaches using pure IMUs do not suffer from occlusion and restricted recording environment and space.
Commercial inertial motion capture systems such as Xsens MVN \cite{Xsens2} use 17 body-worn IMUs which can fully determine the orientations of all bones of a kinematic body model.
However, the large number of sensors are intrusive to the performer and suffer from a long setup time.
Though reducing the number of IMUs can significantly improve user experience, reconstructing human poses from a sparse set of IMUs is a severely underconstrained problem.
Previous works take use of sparse accelerometers \cite{Slyper2008, Tautges2011, Riaz2015} and leverage a prerecorded motion database.
They search in the database for similar accelerations when predicting new poses using a lazy learning strategy \cite{LazyLearning}, and demonstrate that learning with pure accelerometers is feasible.
However, the searching-based methods cannot fully explore the input information.
Due to the fundamental instability of accelerometer measurements and their weak correlations to the poses, the performance of these works is limited.
Also, a trade-off between runtime performance and the database size is inevitable.
In \cite{Schwarz2009}, support vector machine (SVM) and Gaussian processes regression (GPR) are used to categorize the actions and predict full-body poses using sparse orientation measurements.
With the development of the sensing technique, inertial measurement units that measure both accelerations and orientations become common and popular.
Instead of using only accelerations or orientations, recent works leverage both to make full use of IMUs and achieve better accuracy.
Convolutional neural network (CNN) is used in \cite{Hannink2016} to estimate gait parameters from inertial measurements for medical purposes.
The pioneering work, SIP \cite{SIP}, presents a method to solve for the human motions with only 6 IMUs.
As an iterative optimization-based method, it has to operate in an offline manner, making real-time application infeasible.
The state-of-the-art work, DIP \cite{DIP}, proposes to use only 6 IMUs to reconstruct full body poses in real-time.
The authors adopt a bidirectional recurrent neural network (biRNN) \cite{birnn} to directly learn the mapping from IMU measurements to body joint rotations.
DIP achieves satisfying capture quality and efficiency, but there is still space to further improve it.
In our approach, we demonstrate that decomposing this task into multiple stages, i.e. estimating joint positions as an intermediate representation before regressing joint angles, can significantly improve the accuracy and reduce the runtime.
Equally importantly, DIP cannot estimate the global movement of the subject, which is an indispensable part of motion capture.
In this paper, we propose a novel method to estimate the global translation of the performer \textit{without any direct distance measurement}, which is a hybrid of supporting-foot-based deduction and network-based prediction.

\section{Method}
%
\begin{figure*}
    \includegraphics[width=\textwidth]{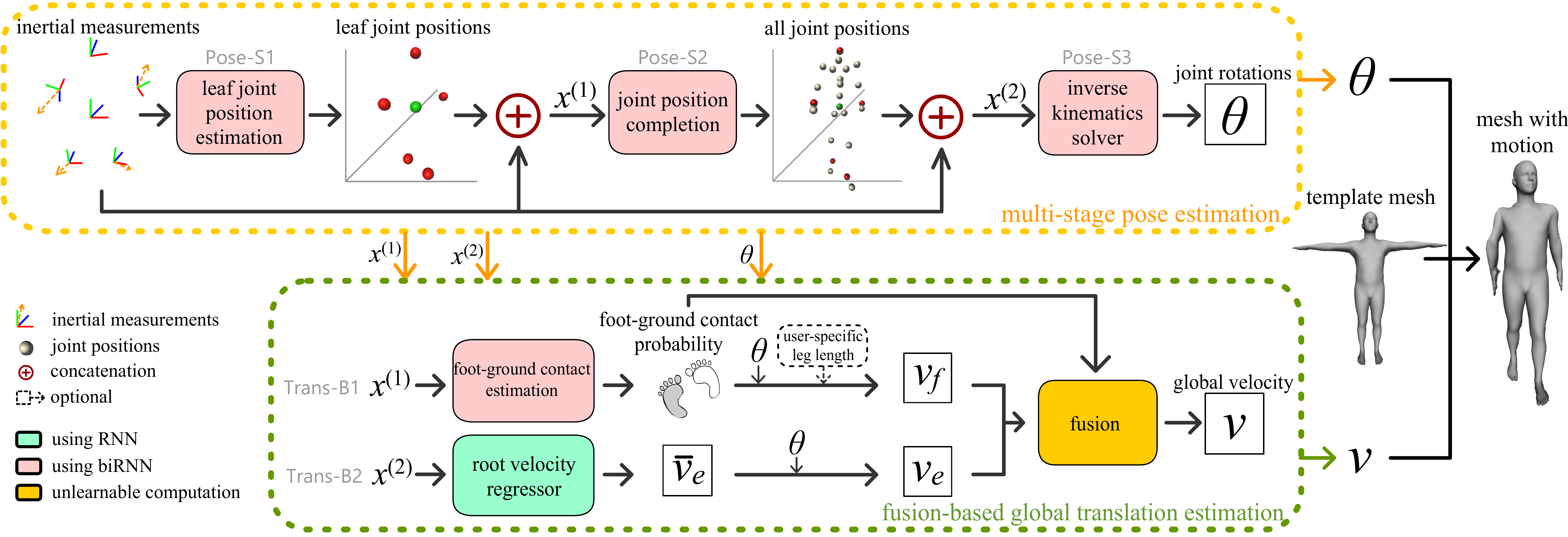}
    \caption{
        Overview of our pipeline.
        We divide our task into 2 subtasks: \textit{multi-stage pose estimation} and \textit{fusion-based global translation estimation}.
        The pose estimation subtask is formulated into 3 stages, where:
        \textit{1)} Pose-S1 estimates 5 leaf joint positions relative to the root from IMU signals, and its output is concatenated to the inertial measurements as $\boldsymbol x^{(1)}$;
        \textit{2)} Pose-S2 estimates all joint positions from $\boldsymbol x^{(1)}$, and the output is again concatenated to the inertial measurements as $\boldsymbol x^{(2)}$;
        \textit{3)} Pose-S3 regresses the joint rotations $\boldsymbol\theta$ from $\boldsymbol x^{(2)}$.
        The translation estimation subtask is addressed by two parallel branches, where:
        \textit{1)} Trans-B1 estimates foot-ground contact probabilities from $\boldsymbol x^{(1)}$ and calculates the root velocity $\boldsymbol v_f$ using forward kinematics assuming that the foot with a higher probability is still;
        \textit{2)} Trans-B2 estimates the local velocity of the root joint $\boldsymbol{\bar v}_e$ in its own coordinate frame from $\boldsymbol x^{(2)}$, which is then transformed to the world space as $\boldsymbol v_e$ according to the root rotation measured by the sensor mounted on the pelvis.
        Such two estimated velocities are then fused to form the final global translation $\boldsymbol v$ according to the foot-ground contact probabilities.
    }
    \Description{Our pipeline flow chart.}
    \label{fig:pipeline}
\end{figure*}
\begin{figure}
    \includegraphics[width=\linewidth]{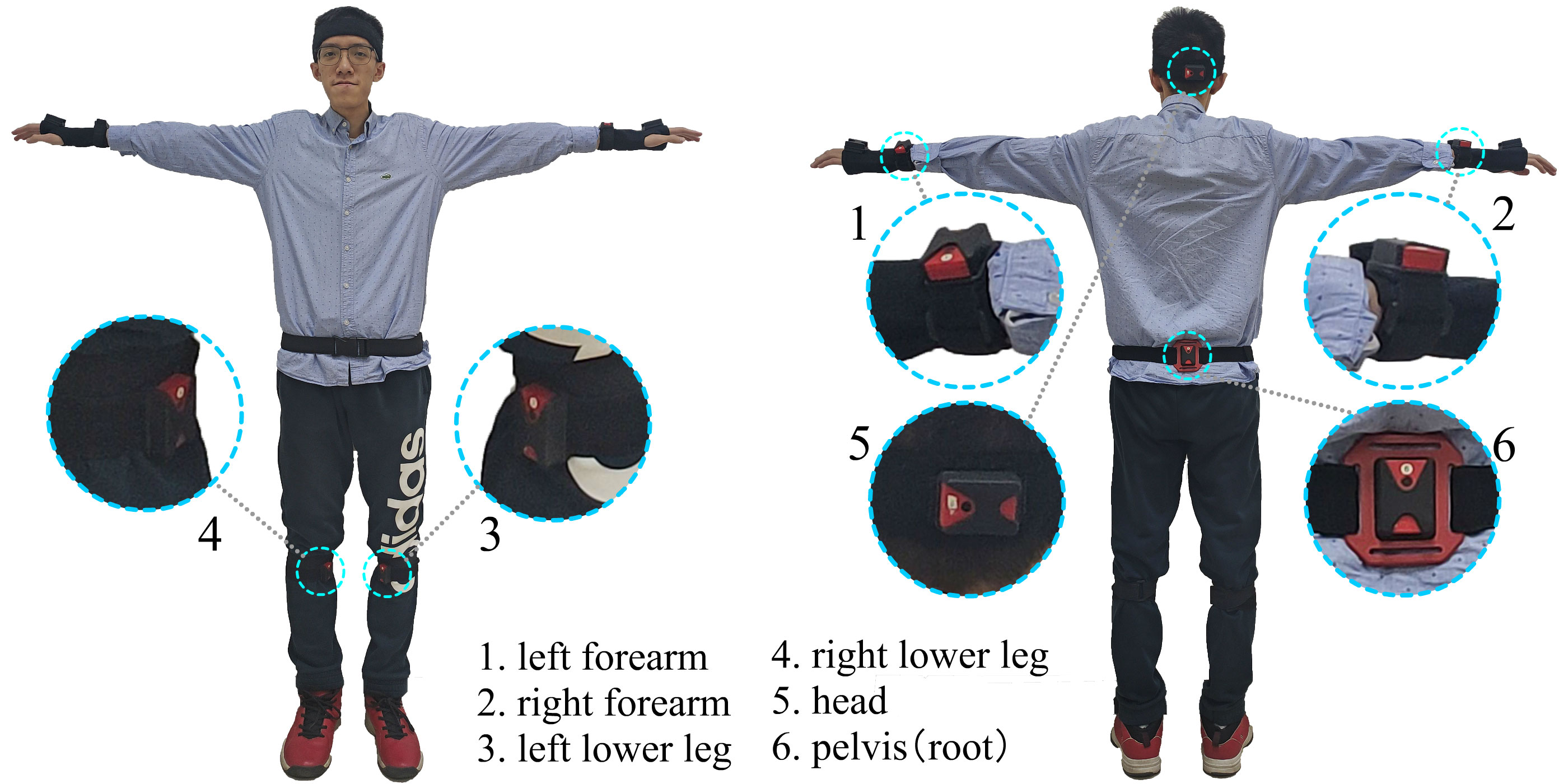}
    \caption{
        IMU placement.
        The 6 IMUs are mounted on the left and right forearm, the left and right lower leg, the head, and the pelvis.
        We require the sensors to be tightly bounded around the joints with arbitrary orientations.
    }
    \Description{Front and back views of a subject wearing 6 IMUs.}
  \label{fig:wearing_imus}
\end{figure}
Our task is to estimate poses and translations of the subject in real-time using 6 IMUs.
As shown in Figure \ref{fig:wearing_imus}, the 6 IMUs are mounted on the pelvis, the left and right lower leg, the left and right forearm, and the head.
In the following, we refer to these joints as \textit{leaf joints} except the pelvis, which is named as \textit{root joint}.
We divide this task into two subtasks: \textit{multi-stage pose estimation} (Section \ref{sec:poseestimation}) and \textit{fusion-based global translation estimation} (Section \ref{sec:transestimation}).
The system is illustrated in Figure \ref{fig:pipeline}.
Detailed structures and hyper-parameters for all networks are presented in Appendix \ref{app:net}. 
%
\subsection{System Input} \label{sec:input}
%
We take the rotation and acceleration measurements of each IMU as the overall input of the system.
We align these measurements into the same reference frame and normalize them to obtain the concatenated input vector as
$\boldsymbol x^{(0)} = [\boldsymbol a_{\mathrm{root}}, \cdots, \boldsymbol a_{\mathrm{rarm}}, \boldsymbol R_{\mathrm{root}}, \cdots, \boldsymbol R_{\mathrm{rarm}}] \in \mathbb{R}^{72}$
where $\boldsymbol a \in \mathbb{R}^3$ is the acceleration and $\boldsymbol R \in \mathbb{R}^{3 \times 3}$ is the rotation matrix.
We use $\boldsymbol x^{(0)}(t)$ to refer to the measurements of the $t$th frame, and the superscript $^{(0)}$ means it is the overall input.
Please refer to Appendix \ref{app:imu} for more details on the sensor preprocessing.
%
\subsection{Multi-stage Pose Estimation} \label{sec:poseestimation}
%
In this section, we introduce our multi-stage method to estimate body poses, i.e. the rotation of each joint, from sparse IMU measurements.
This task is very difficult due to the ambiguity of the mapping from extremely sparse inertial data to full body joint angles.
The key to address this challenging task is the use of \textit{1)} the prior knowledge on human poses and \textit{2)} the temporal information.
Because of the diversity of human motions and the anatomic restrictions, the pose prior is hard to model or learn.
While the previous work, DIP \cite{DIP}, proposes to adopt a bidirectional recurrent neural network (biRNN) to learn the direct mapping from IMU measurements to joint rotations, we demonstrate that using joint positions as an intermediate representation is significantly helpful for the model to learn the complex motion prior.
More specifically, we first predict joint positions as an intermediate task in Pose-S1 (short for \textit{pose estimation stage 1}) and Pose-S2 (Section \ref{sec:stage12}), and then solve for the joint angles in Pose-S3 (Section \ref{sec:stage3}).
Further, to fully utilize the temporal information, we adopt biRNNs to leverage the past and future frames.
As shown in Section \ref{sec:comparisons}, our approach achieves superior capture quality and runtime performance.
%
\subsubsection{Pose-S1\&S2: estimating joint positions} \label{sec:stage12}
%
In Pose-S1\&S2, we estimate joint positions from IMU measurements.
We explicitly separate this task into two different stages to take the hierarchy of the human kinematic tree into consideration.
Specifically, we divide the human joints into two sets: the leaf joints and the non-leaf joints.
The leaf joints are mounted with sensors and have direct measurements.
Usually, they have larger movements.
The inner non-leaf joints are without any direct inertial data and have relatively smaller motions.
Due to the correlation between body joints, the information of the leaf joints is helpful for predicting the coordinates of inner ones.
We first regress the positions of the leaf joints in Pose-S1, and then localize the other joints in Pose-S2.
%
\par
%
The input to Pose-S1 is the inertial measurement vector $\boldsymbol x^{(0)}(t)$, and the output is the root-relative positions of the five leaf joints $\boldsymbol p_{\mathrm{leaf}}(t) = [\boldsymbol p_{\mathrm{lleg}}(t), \cdots, \boldsymbol p_{\mathrm{rarm}}(t)] \in \mathbb{R}^{15}$.
We use a standard biRNN \cite{birnn} with Long Short-Term Memory (LSTM) cells \cite{lstm} to learn the mapping from IMU measurements to leaf joint positions.
Due to the very sparse measurements, it is common that the IMUs show identical signals while the subject is performing different animations.
For example, when sitting still or standing still, the acceleration and orientation measurements of all the IMUs are almost the same.
In such cases, the temporal information is the key to resolve this ambiguity, thus using RNNs is a natural choice.
We choose to use biRNNs instead of vanilla RNNs because the future information is also of great help in this task, as stated in DIP.
The loss function used to train the network is defined as:
\begin{equation}
	\mathcal{L}_{\mathrm{S1}} = \|\boldsymbol p_{\mathrm{leaf}}(t) - \boldsymbol p_{\mathrm{leaf}}^{\mathrm{GT}}(t)\|_2^2 \mathrm{,}
\end{equation}
where superscript $^{\mathrm{GT}}$ denotes the ground truth.
%
\par
%
The input of Pose-S2 is the concatenation of Pose-S1's output and the inertial measurements: $\boldsymbol x^{(1)}(t) = [\boldsymbol p_{\mathrm{leaf}}(t), \boldsymbol x^{(0)}(t)] \in \mathbb{R}^{87}$.
Based on the input information, Pose-S2 outputs the root-relative coordinates of all joints: $\boldsymbol p_{\mathrm{all}}(t) = [\boldsymbol p_j(t) | j = 1,2,\cdots,J-1] \in \mathbb{R}^{3(J-1)}$, where $J$ is the number of joints in the human kinematic tree.
Similar to Pose-S1, we use a biRNN for Pose-S2 with L2 loss:
\begin{equation}
	\mathcal{L}_{\mathrm{S2}} = \|\boldsymbol p_{\mathrm{all}}(t) - \boldsymbol p_{\mathrm{all}}^{\mathrm{GT}}(t)\|_2^2 \mathrm{.}
\end{equation}
%
\subsubsection{Pose-S3: estimating joint rotations} \label{sec:stage3}
%
In Pose-S3, we estimate joint rotations from joint positions.
We concatenate the joint coordinates and the inertial measurements as $\boldsymbol x^{(2)}(t)=[\boldsymbol p_{\mathrm{all}}(t), \boldsymbol x^{(0)}(t)] \in \mathbb{R}^{3(J-1)+72}$, which is the input of Pose-S3.
The input vector is fed into a biRNN, which predicts the rotations of all non-root joints relative to the root in the 6D representation \cite{6D}: $\boldsymbol R_{\mathrm{all}}^{(\mathrm{6D})}(t)=[\boldsymbol R_j^{(\mathrm{6D})}(t) | j = 1, 2, \cdots, J-1] \in \mathbb{R}^{6(J-1)}$.
We choose the 6D rotation representation over other common representations such as quaternions as the output for better continuity, as demonstrated in \cite{6D}.
We should note that the IMU measurements are fed into the networks as rotation matrices, while the only use of the 6D representation is in the output of Pose-S3.
The loss function is defined as:
\begin{equation}
	\mathcal{L}_{\mathrm{S3}} = \|\boldsymbol R_{\mathrm{all}}^{(\mathrm{6D})}(t) - \boldsymbol R_{\mathrm{all}}^{\mathrm{GT},(\mathrm{6D})}(t)\|_2^2 \mathrm{.}
\end{equation}
The rotation of the root joint $\boldsymbol R_{\mathrm{root}}$ is directly measured by the sensor placed on the pelvis.
Combining all these rotations, we convert them to the rotation matrix formulation and obtain the full body pose as $\boldsymbol \theta = [\boldsymbol R_j(t) | j = 0,1,\cdots,J] \in \mathbb{R}^{9J}$.
%
\subsection{Fusion-based Global Translation Estimation} \label{sec:transestimation}
%
In this section, we explain our method to estimate global translations from the IMU measurements and the estimated body poses.
This task is even more challenging due to the lack of direct distance measurements, and the acceleration measurements are too noisy to be used directly \cite{SIP}.
Previous works address this task by introducing additional vision inputs \cite{Malleson2017, Malleson2019, Henschel2020, Andrews2016} or distance measurements \cite{Vlasic2007, Liu2011}, which increase the complexity of the system.
While the work of SIP \cite{SIP} estimates global translations from IMUs only, it has to run in an offline manner.
To our best knowledge, we are the first method that addresses the task of real-time prediction of global translations from sparse IMUs.
To do so, we need to estimate the per-frame velocity of the root joint, i.e. the translation between the current frame and the previous frame.
We propose a fusion-based approach that comprises two parallel branches.
In Trans-B1 (short for \textit{translation estimation branch 1}) (Section \ref{sec:branch1}), we infer the root velocity based on the sequential pose estimation, in combination with the prediction of foot-ground contact probabilities.
In Trans-B2 (Section \ref{sec:branch2}), we use an RNN to predict the root velocity from joint positions and inertial measurements.
These two branches run in a parallel style, and the final estimation is a fusion of the two branches based on the foot state (Section \ref{sec:merge}).
The intuition behind is that by assuming the foot on the ground is not moving, the per-frame velocity can be deduced from the motion of the subject.
However, this estimation is not totally accurate and inevitably fails when both feet are not on the ground, e.g., jumping or running.
Therefore, a neural network is used here for complementary estimation.
We demonstrate in Section \ref{sec:evaluations} that such a hybrid approach gives more accurate results than using either branch alone.
%
\subsubsection{Trans-B1: supporting-foot-based velocity estimation} \label{sec:branch1}
%
In Trans-B1, we estimate the velocity of the root joint based on the estimation of body poses.
First, we use a biRNN to estimate the foot-ground contact probability for each foot, i.e. the likelihood that the foot is on the ground, formulated as $\boldsymbol s_{\mathrm{foot}} = [s_{\mathrm{lfoot}}, s_{\mathrm{rfoot}}] \in \mathbb{R}^2$.
The input of this network is the leaf joint positions and the inertial measurements, $\boldsymbol x^{\mathrm{(1)}}$.
We assume that the foot with a higher on-ground probability is not moving between two adjacent frames, referred to as the supporting foot.
The corresponding probability is denoted as $s = \max\{s_{\mathrm{lfoot}}, s_{\mathrm{rfoot}}\}$.
Then, we apply the estimated pose parameters $\boldsymbol \theta$ on the kinematic model with optional user-specific leg lengths $\boldsymbol l$, and the root velocity is essentially the coordinate difference of the supporting foot between two consecutive frames, denoted as $\boldsymbol v_f$:
\begin{equation}
    \boldsymbol v_f(t) = \mathrm{FK}(\boldsymbol \theta(t - 1); \boldsymbol l) - \mathrm{FK}(\boldsymbol \theta(t); \boldsymbol l)  \mathrm{,}
\end{equation}
where $\mathrm{FK}(\cdot)$ is the forward kinematics function that calculates the position of the supporting foot from the pose parameters.
To train the model, we use a cross-entropy loss defined as:
\begin{equation}
	\begin{split}
	\mathcal{L}_{\mathrm{B1}} = &-s_{\mathrm{lfoot}}^{\mathrm{GT}} \log s_{\mathrm{lfoot}} - (1-s_{\mathrm{lfoot}}^{\mathrm{GT}}) \log (1-s_{\mathrm{lfoot}})\\
	                            &-s_{\mathrm{rfoot}}^{\mathrm{GT}} \log s_{\mathrm{rfoot}} - (1-s_{\mathrm{rfoot}}^{\mathrm{GT}}) \log (1-s_{\mathrm{rfoot}}) \mathrm{.}
	\end{split}
\end{equation}
%
\subsubsection{Trans-B2: network-based velocity estimation} \label{sec:branch2}
%
While the pose-based method in Trans-B1 is straightforward, it is substantially incapable of the cases where both feet are off the ground simultaneously.
Further, the errors in the pose estimation will also affect the accuracy.
To this end, we additionally adopt a neural network to estimate the velocity based on the predicted joint positions and the IMU measurements, $\boldsymbol x^{\mathrm{(2)}}$, in parallel to Trans-B1.
Instead of using a biRNN as in other steps, here we choose to use an RNN.
This is because we find that estimating per-frame velocity accurately requires more previous frames, and a biRNN with a sufficiently large window size cannot satisfy our runtime requirements (see the experiments in Section \ref{sec:evaluations}).
The output of the RNN is the velocity $\boldsymbol{\bar v}_e$ in the coordinate system of the root joint, and we transform it to the global system using the root rotation $\boldsymbol{R}_{\mathrm{root}}$ in $\boldsymbol \theta$ as:
\begin{equation}
	\boldsymbol v_e = \boldsymbol{R}_{\mathrm{root}} \boldsymbol{\bar v}_e \mathrm{.}
\end{equation}
The loss function is defined as:
\begin{equation}\label{equ:accloss}
	\mathcal{L}_{\mathrm{B2}} = \mathcal{L}_{\mathrm{vel}}(1) + \mathcal{L}_{\mathrm{vel}}(3) + \mathcal{L}_{\mathrm{vel}}(9) + \mathcal{L}_{\mathrm{vel}}(27) \mathrm{,}
\end{equation}
where:
\begin{equation}
	\mathcal{L}_{\mathrm{vel}}(n) = \sum_{m=0}^{\lfloor T/n \rfloor - 1} \left\|\sum_{t=m}^{mn+n-1} (\boldsymbol {\bar v}_e(t) - \boldsymbol {\bar v}^{\mathrm{GT}}_e(t)) \right\|_2^2 \mathrm{.}
\end{equation}
Here, $T$ is the total number of frames in the training sequence.
This loss measures the differences between the predicted and the ground truth translations in every consecutive 1, 3, 9, and 27 frames.
We find that using $\mathcal{L}_{\mathrm{vel}}(1)$ alone makes the network only focus on neighboring frames, resulting in unstable estimation and large errors.
On the contrary, the supervision on larger frame ranges helps to reduce the accumulative error.
%
\subsubsection{Fusing two branches} \label{sec:merge}
%
Lastly, we fuse the estimated velocity $\boldsymbol v_f$ and $\boldsymbol v_e$ to get the final global translation $\boldsymbol v$ based on the foot-ground contact likelihood.
We set an upper threshold $\overline s$ and a lower threshold $\underline s$ for the predicted foot-ground contact probability $s$.
The output global velocity is then computed as:
\begin{equation}
	\boldsymbol v = \left\{
		\begin{array}{lll}
			\boldsymbol v_e & & {0 \le s < \underline s}\\
			\frac{s - \overline s}{\underline s - \overline s}\boldsymbol v_e + \frac{s - \underline s}{\overline s - \underline s}\boldsymbol v_f & & {\underline s \le s < \overline s}\\
			\boldsymbol v_f & & {\overline s \le s \le 1}
		\end{array}
	\right. \mathrm{.}
\end{equation}
We empirically set $\underline s = 0.5$ and $\overline s = 0.9$ in our model.
The linear interpolation provides a smooth and automatic transition between the two branches.

\section{Implementation}
%
In this section, we provide more details on our implementation.
Specifically, we introduce the kinematic model in Section \ref{sec:kinematicmodel}, the training data in Section \ref{sec:trainingdata}, and other related details in Section \ref{sec:otherdetails}.
%
\subsection{Kinematic Model} \label{sec:kinematicmodel}
%
We use the SMPL \cite{SMPL} skeleton as our kinematic model, and use the corresponding mesh for visualization.
The model is defined as:
\begin{equation}
	M(\boldsymbol{\theta}) = W(\boldsymbol{\bar{\rm T}}, \boldsymbol{J}, \boldsymbol{\theta}, \mathcal{W}) \mathrm{,} \label{equ:smpl}
\end{equation}
where $\boldsymbol{\bar{\rm T}}$ is the template mesh in the rest pose, $\boldsymbol{J}$ is 24 body joints, $\boldsymbol{\theta}$ is the pose parameters in terms of joint angles, $\mathcal{W}$ is the blend weights, and $W(\cdot)$ is the linear blend skinning function.
As the body shape is not our concern, we leave out the shape- and pose-blendshape terms in the original paper.
We should note that the IMU measurements do not contain any information about the subject's body shape, and the pose estimation task is also shape-agnostic.
However, in Trans-B1, leg lengths will directly affect the prediction of global translations which are deduced from foot movements.
Here, we assume that the leg lengths of the subject can be measured in advance, and the global translations are calculated based on this measurement.
Otherwise, we use the mean SMPL model, which will give a less accurate estimation due to the disparity in leg lengths, but still being plausible.
%
\subsection{Training Data} \label{sec:trainingdata}
%
In our implementation, each network requires different types of data and is trained individually.
The relevant datasets include:
\textit{1)} DIP-IMU \cite{DIP}, which consists of IMU measurements and pose parameters for around 90 minutes of motions performed by 10 subjects wearing 17 IMUs;
\textit{2)} TotalCapture \cite{TotalCapture}, which consists of IMU measurements, pose parameters, and global translations for around 50 minutes of motions performed by 5 subjects wearing 13 IMUs;
\textit{3)} AMASS \cite{AMASS}, which is a composition of existing motion capture (mocap) datasets and contains pose parameters and global translations for more than 40 hours of motions performed by over 300 subjects.
The overview of the datasets is shown in Table \ref{tab:dataset}.
As TotalCapture is relatively small, we use it only for evaluation as an examination for cross-dataset generalization. 
In the following, we introduce how to use the training data for each subtask, i.e. pose estimation (Section \ref{sec:posedata}) and global translation estimation (Section \ref{sec:transdata}).
\begin{table}[]
	\caption{
	    Dataset overview. 
	    We use DIP-IMU \cite{DIP}, TotalCapture \cite{TotalCapture}, and AMASS \cite{AMASS} dataset for our training and evaluation. 
	    The table shows pose parameters, IMU measurements, global translations, foot-ground contact states, and the total length in minutes for each dataset. 
	    "Y" means that the dataset contains such information. 
	    "N" means that the dataset does not contain such information. 
	    "S" means that the data is synthesized from other information.
	}
	\label{tab:dataset}
	\begin{minipage}{\columnwidth}
	\begin{center}
	\begin{tabular}{cccccc}
		\toprule
		Dataset & Pose & IMU & Translation & Contact & Minutes \\
		\midrule
		DIP-IMU\footnote{Some training sequences that have too many "nan" IMU measurements are discarded.}  & Y   & Y   & N   & N   & 80    \\
		TotalCapture & Y\footnote{The ground truth SMPL pose parameters are provided by DIP \cite{DIP}.}          & Y   & Y   & S   & 49    \\
		AMASS\footnote{We select and sample the original AMASS motions \cite{AMASS} in 60 fps.}             & Y   & S   & Y   & S   & 1217  \\
		\bottomrule
	\end{tabular}
	\end{center}
	\end{minipage}
\end{table}
%
\subsubsection{Training data for pose estimation} \label{sec:posedata}
%
To train the networks in Pose-S1 and Pose-S2, we need sequential IMU measurements (including rotations and accelerations) and root-relative joint coordinates.
While DIP-IMU contains such data, it is not diverse enough to train a model with sufficient generalization ability.
Hence, we synthesize inertial data for AMASS dataset which is much larger and contains more variations.
To do so, we place virtual IMUs on the corresponding vertices of the SMPL mesh, and then the sequential positions and rotations of each IMU can be inferred from the pose parameters using Equation \ref{equ:smpl}.
We use the following method to synthesize the acceleration measurements:
\begin{equation} \label{equ:syn}
	\boldsymbol a_i(t) = \frac{\boldsymbol x_i(t-n) + \boldsymbol x_i(t+n) - 2\boldsymbol x_i(t)}{(n\Delta t)^2}, i = 1,2,\cdots,6 \mathrm{,}
\end{equation}
where $\boldsymbol x_i(t)$ denotes the coordinate of the $i$th IMU at frame $t$, $\boldsymbol a_i(t) \in \mathbb{R}^3$ is the acceleration of IMU $i$ at frame $t$, and $\Delta t$ is the time interval between two consecutive frames.
To cope with jitters in the data, we do not compute the accelerations based on adjacent frames (i.e. $n = 1$), but use relatively apart frames for smoother accelerations by setting $n = 4$.
Please refer to Appendix \ref{app:syn} for more details.
We synthesize a subset of AMASS dataset which consists of about 1217 minutes of motions in 60 fps, which sufficiently cover the motion variations.
To make Pose-S2 robust to the prediction errors of leaf-joint positions, during training, we further add Gaussian noise to the leaf-joint positions with $\sigma = 0.04$.
We use DIP-IMU and synthetic AMASS as training datasets for the pose estimation task, and leave TotalCapture for evaluation.
To train Pose-S3, we need inertial measurements and mocap data in the form of joint angles.
We again use DIP-IMU and synthetic AMASS dataset for Pose-S3 with additional Gaussian noise added to the joint positions, whose standard deviation is empirically set to $\sigma = 0.025$.
%
\subsubsection{Training data for translation estimation} \label{sec:transdata}
%
In Trans-B1, a biRNN is used to predict foot-ground contact probabilities, thus we need binary annotations for foot-ground contact states.
To generate such data, we apply the 60-fps pose parameters and root translations on the SMPL model to obtain the coordinates of both feet, denoted as $\boldsymbol x_{\mathrm{lfoot}}(t)$ and $\boldsymbol x_{\mathrm{rfoot}}(t)$ for frame $t$.
When the movement of one foot between two consecutive frames is less than a threshold $u$, we mark it as contacting the ground.
We empirically set $u = 0.008$ meters.
In this way, we automatically label AMASS dataset using:
\begin{equation}
	s_{\mathrm{lfoot}}^{\mathrm{GT}}(t) = \left\{
	\begin{array}{lll}
		1 & & \mathrm{if}\,\, \|\boldsymbol x_{\mathrm{lfoot}}(t) - \boldsymbol x_{\mathrm{lfoot}}(t-1)\|_2 < u\\
		0 & & \mathrm{otherwise}
	\end{array}
	\right.\mathrm{,}
\end{equation}
\begin{equation}
	s_{\mathrm{rfoot}}^{\mathrm{GT}}(t) = \left\{
	\begin{array}{lll}
		1 & & \mathrm{if}\,\, \|\boldsymbol x_{\mathrm{rfoot}}(t) - \boldsymbol x_{\mathrm{rfoot}}(t-1)\|_2 < u\\
		0 & & \mathrm{otherwise}
	\end{array}
	\right.\mathrm{,}
\end{equation}
where $s_{\mathrm{lfoot}}^{\mathrm{GT}}(t)$ and $s_{\mathrm{rfoot}}^{\mathrm{GT}}(t)$ are the foot-ground contact state labels for frame $t$.
For better robustness, we additionally apply Gaussian noise to the input of Trans-B1 during training, with the standard deviation set to $\sigma = 0.04$.
Since DIP-IMU does not contain global translations, we only use synthetic 60-fps AMASS as the training dataset and leave TotalCapture for evaluation.
%
\par
%
In Trans-B2, an RNN is used to predict the velocity of the root joint in its own coordinate system.
Since the root positions are already provided in AMASS dataset, we compute the velocity and convert it from the global space into the root space using:
\begin{equation}
	\boldsymbol {\bar v}_e^{\mathrm{GT}}(t) = (\boldsymbol R_{\mathrm{root}}^{\mathrm{GT}}(t))^{-1}(\boldsymbol {x}_{\mathrm{root}}^{\mathrm{GT}}(t)-\boldsymbol {x}_{\mathrm{root}}^{\mathrm{GT}}(t-1)) \mathrm{,}
\end{equation}
where $\boldsymbol R_{\mathrm{root}}^{\mathrm{GT}}(t)$ is the ground truth root rotation at frame $t$ and $\boldsymbol {x}_{\mathrm{root}}^{\mathrm{GT}}(t)$ is the ground truth root position in world space at frame $t$.
Note that the frame rate is fixed to be 60 fps for training and testing, and the \textit{velocity} is defined as the translation between two consecutive 60-fps frames in this paper.
We intend to use Trans-B2 mainly for the cases where both feet are off the ground, thus we only use such kind of sequences from AMASS to construct the training data. 
Specifically, we run the well-trained Trans-B1 network and collect the sequence clips where the minimum foot-ground contact probability is lower than $\overline s$, i.e. $\min_{t \in \mathcal{F}}s(t) < \overline s$, where $\mathcal{F}$ is the set containing all frames of the fragment and $|\mathcal F| \le 300$.
During training, Gaussian noise of $\sigma = 0.025$ is added to the input of Trans-B2.
%
\subsection{Other Details} \label{sec:otherdetails}
%
All training and evaluation processes run on a computer with an Intel(R) Core(TM) i7-8700 CPU and an NVIDIA GTX1080Ti graphics card.
The live demo runs on a laptop with an Intel(R) Core(TM) i7-10750H CPU and an NVIDIA RTX2080 Super graphics card.
The model is implemented using PyTorch 1.7.0 with CUDA 10.1.
The front end of our live demo is implemented using Unity3D, and we use Noitom\footnote{\url{https://www.noitom.com/}} Legacy IMU sensors to collect our own data.
We separately train each network with the batch size of 256 using an Adam \cite{Adam} optimizer with a learning rate $\mathrm{lr}=10^{-3}$.
We follow DIP to train the models for the pose estimation task using synthetic AMASS first and fine-tune them on DIP-IMU which contains real IMU measurements.
To avoid the vertical drift due to the error accumulation in the estimation of translations, we add a gravity velocity $v_G = 0.018$ to the Trans-B1 output $\boldsymbol v_f$ to pull the body down.

\section{Experiments}
%
In this section, we first introduce the data and metrics used in our experiments in Section \ref{sec:data}.
Using the data and metrics, we compare our method with previous methods qualitatively and quantitatively in Section \ref{sec:comparisons}.
Next, we evaluate our important design choices and key technical components by ablative studies in Section \ref{sec:evaluations}.
Finally, to further demonstrate the power of our technique, we show various real-time results containing strong occlusion, wide range motion space, dark and outdoor environment, close interaction, and multiple users in Section \ref{sec:livedemo}.
Following DIP \cite{DIP}, we adopt two settings of inference: the \textit{offline} setting where the full sequence is available at the test time, and the \textit{online} (\textit{real-time}) setting where our approach accesses 20 past frames, 1 current frame, and 5 future frames in a window sliding manner, with a tolerable latency of $83 \mathrm{ms}$.
Please note that all the results are estimated from the real IMU measurements, and no temporal filter is used in our method.
Following DIP, we do not give rotation freedoms to the wrists and ankles because we have no observation to solve them.
%
\subsection{Data and Metrics} \label{sec:data}
%
The pose evaluations are conducted on the test split of DIP-IMU \cite{DIP} and TotalCapture \cite{TotalCapture} dataset.
The global translation evaluations are conducted on TotalCapture alone, as DIP-IMU does not contain global movements.
We use the following metrics for quantitative evaluations of poses:
\textit{1)} {\it SIP error} measures the mean global rotation error of upper arms and upper legs in degrees;
\textit{2)} {\it angular error} measures the mean global rotation error of all body joints in degrees;
\textit{3)} {\it positional error} measures the mean Euclidean distance error of all estimated joints in centimeters with the root joint (Spine) aligned;
\textit{4)} {\it mesh error} measures the mean Euclidean distance error of all vertices of the estimated body mesh also with the root joint (Spine) aligned.
The vertex coordinates are calculated by applying the pose parameters to the SMPL \cite{SMPL} body model with the mean shape using Equation \ref{equ:smpl}.
Note that the errors in twisting (rotations around the bone) cannot be measured by positional error, but will be reflected in mesh error.
\textit{5)} {\it Jitter error} measures the average {\it jerk} of all body joints in the predicted motion.
Jerk is the third derivative of position with respect to time and reflects the smoothness and naturalness of the motion \cite{Jerk}.
A smaller average jerk means a smoother and more natural animation.
%
\subsection{Comparisons} \label{sec:comparisons}
%
Quantitative and qualitative comparisons with DIP \cite{DIP} and SIP/SOP (SOP is a simplified version of SIP that only uses orientation measurements) \cite{SIP} are conducted and the results are shown in this section.
We should note that our test data is slightly different from \cite{DIP} for both DIP-IMU and TotalCapture, resulting in some inconsistency with the values reported by DIP.
More specifically, the DIP-IMU dataset contains both raw and calibrated data, and the results in the DIP paper are based on the latter. 
However, the calibrated data removes the root inertial measurements which are required by our method to estimate the global motions, so we have to perform the comparisons on the raw data. 
As for the TotalCapture dataset, the ground truth SMPL pose parameters are acquired from the DIP authors.
To evaluate DIP on the test data, we use the DIP model and the code released by the authors.
No temporal filtering technique is applied.
%
\subsubsection{Quantitative comparisons} \label{sec:quantitative}
%
\begin{table*}[ht]
	\caption{
	    Offline comparison results for body poses. 
	    We compare our method with SIP/SOP \cite{SIP} and DIP \cite{DIP} on TotalCapture \cite{TotalCapture} and DIP-IMU \cite{DIP} dataset. 
	    The mean and standard deviation (in parentheses) of the SIP error, angular error, positional error, mesh error, and jitter error are reported.
	}
	\label{tab:offpose}
	\resizebox{\linewidth}{!}{
	\begin{tabular}{ccccccccccc}
		\toprule
		& \multicolumn{5}{c}{TotalCapture} & \multicolumn{5}{c}{DIP-IMU} \\
		\cmidrule(lr){2-6}\cmidrule(lr){7-11}
		& SIP Err ($\rm{deg}$) & Ang Err ($\rm{deg}$) & Pos Err ($\rm{cm}$) & Mesh Err ($\rm{cm}$) & Jitter ($10^2\rm{m}/\rm{s}^3$) &
		  SIP Err ($\rm{deg}$) & Ang Err ($\rm{deg}$) & Pos Err ($\rm{cm}$) & Mesh Err ($\rm{cm}$) & Jitter ($10^2\rm{m}/\rm{s}^3$) \\
		\midrule
		SOP  & 23.09 ($\pm$12.37) & 17.14 ($\pm$8.54) &  9.24 ($\pm$5.33) & 10.58 ($\pm$6.04) &  8.17 ($\pm$13.55) &
		       24.56 ($\pm$12.75) &  9.83 ($\pm$5.21) &  8.17 ($\pm$4.74) &  9.32 ($\pm$5.27) &  5.66 ($\pm$9.49) \\
		SIP  & 18.54 ($\pm$9.67) & 14.84 ($\pm$7.26) &  7.65 ($\pm$4.32) &  8.60 ($\pm$4.83) &  8.27 ($\pm$17.36) &
		       21.02 ($\pm$9.61) &  8.77 ($\pm$4.38) &  6.66 ($\pm$3.33) &  7.71 ($\pm$3.80) &  3.86 ($\pm$6.32) \\
		DIP  & 18.79 ($\pm$11.85) & 17.77 ($\pm$9.51) &  9.61 ($\pm$5.76) & 11.34 ($\pm$6.45) & 28.86 ($\pm$29.18) &
		       16.36 ($\pm$8.60) & 14.41 ($\pm$7.90) &  6.98 ($\pm$3.89) &  8.56 ($\pm$4.65) & 23.37 ($\pm$23.84) \\
		Ours & \textbf{14.95 ($\pm$6.90)} & \textbf{12.26 ($\pm$5.59)} & \textbf{5.57 ($\pm$3.09)} & \textbf{6.36 ($\pm$3.47)} & \textbf{1.57 ($\pm$2.93)} &
		       \textbf{13.97 ($\pm$6.77)} & \textbf{ 7.62 ($\pm$4.01)} & \textbf{4.90 ($\pm$2.75)} & \textbf{5.83 ($\pm$3.21)} & \textbf{1.19 ($\pm$1.76)} \\
		\bottomrule
	\end{tabular}}
\end{table*}
\begin{table*}[ht]
	\caption{
	    Online comparison results for body poses. 
	    We compare our method with DIP \cite{DIP} in the real-time setting on TotalCapture \cite{TotalCapture} and DIP-IMU \cite{DIP} dataset.
	}
	\label{tab:onpose}
	\resizebox{\linewidth}{!}{
	\begin{tabular}{ccccccccccc}
		\toprule
		& \multicolumn{5}{c}{TotalCapture} & \multicolumn{5}{c}{DIP-IMU} \\
		\cmidrule(lr){2-6}\cmidrule(lr){7-11}
		& SIP Err ($\rm{deg}$) & Ang Err ($\rm{deg}$) & Pos Err ($\rm{cm}$) & Mesh Err ($\rm{cm}$) & Jitter ($10^2\rm{m}/\rm{s}^3$) &
		SIP Err ($\rm{deg}$) & Ang Err ($\rm{deg}$) & Pos Err ($\rm{cm}$) & Mesh Err ($\rm{cm}$) & Jitter ($10^2\rm{m}/\rm{s}^3$) \\
		\midrule
		DIP  & 18.93 ($\pm$12.44) & 17.50 ($\pm$10.10) & 9.57 ($\pm$5.95) & 11.40 ($\pm$6.87) & 35.94 ($\pm$34.45) &
			   17.10 ($\pm$9.59) & 15.16 ($\pm$8.53) & 7.33 ($\pm$4.23) & 8.96 ($\pm$5.01) & 30.13 ($\pm$28.76) \\
		Ours & \textbf{16.69 ($\pm$8.79)} & \textbf{12.93 ($\pm$ 6.15)} & \textbf{6.61 ($\pm$ 3.93)} & \textbf{7.49 ($\pm$ 4.35)} & \textbf{9.44 ($\pm$13.57)} &
		       \textbf{16.68 ($\pm$8.68)} & \textbf{8.85 ($\pm$ 4.82)} & \textbf{5.95 ($\pm$ 3.65)} & \textbf{7.09 ($\pm$ 4.24)} & \textbf{6.11  ($\pm$7.92)} \\
		\bottomrule
	\end{tabular}}
\end{table*}
The quantitative comparisons are shown in Table \ref{tab:offpose} for the offline setting and Table \ref{tab:onpose} for the online setting.
We report the mean and standard deviation for each metric in comparison with DIP (both offline and online) and SIP/SOP (only offline).
As shown in the tables, our method outperforms all previous works in all metrics.
We attribute our superiority to the multi-stage structure that first estimates joint coordinates in the positional space and then solves for the angles in the rotational space.
We argue that human poses are easier to estimate in the joint-coordinate space, and with the help of the coordinates, joint rotations can be better estimated.
A fly in the ointment is that we can see an accuracy gap between online and offline settings.
The reason is that the offline setting uses much more temporal information to solve the pose of the current frame, while the online setting does not do this to avoid noticeable delay.
Since our approach fully exploits such temporal information to resolve the motion ambiguities, there is an inevitable accuracy reduction when switching from the offline setting to the online setting.
Nevertheless, we still achieve the state-of-the-art online capture quality which is visually pleasing as shown in the qualitative results and the live demos.
%
\subsubsection{Qualitative comparisons}
%
\begin{figure}
	\includegraphics[width=\linewidth]{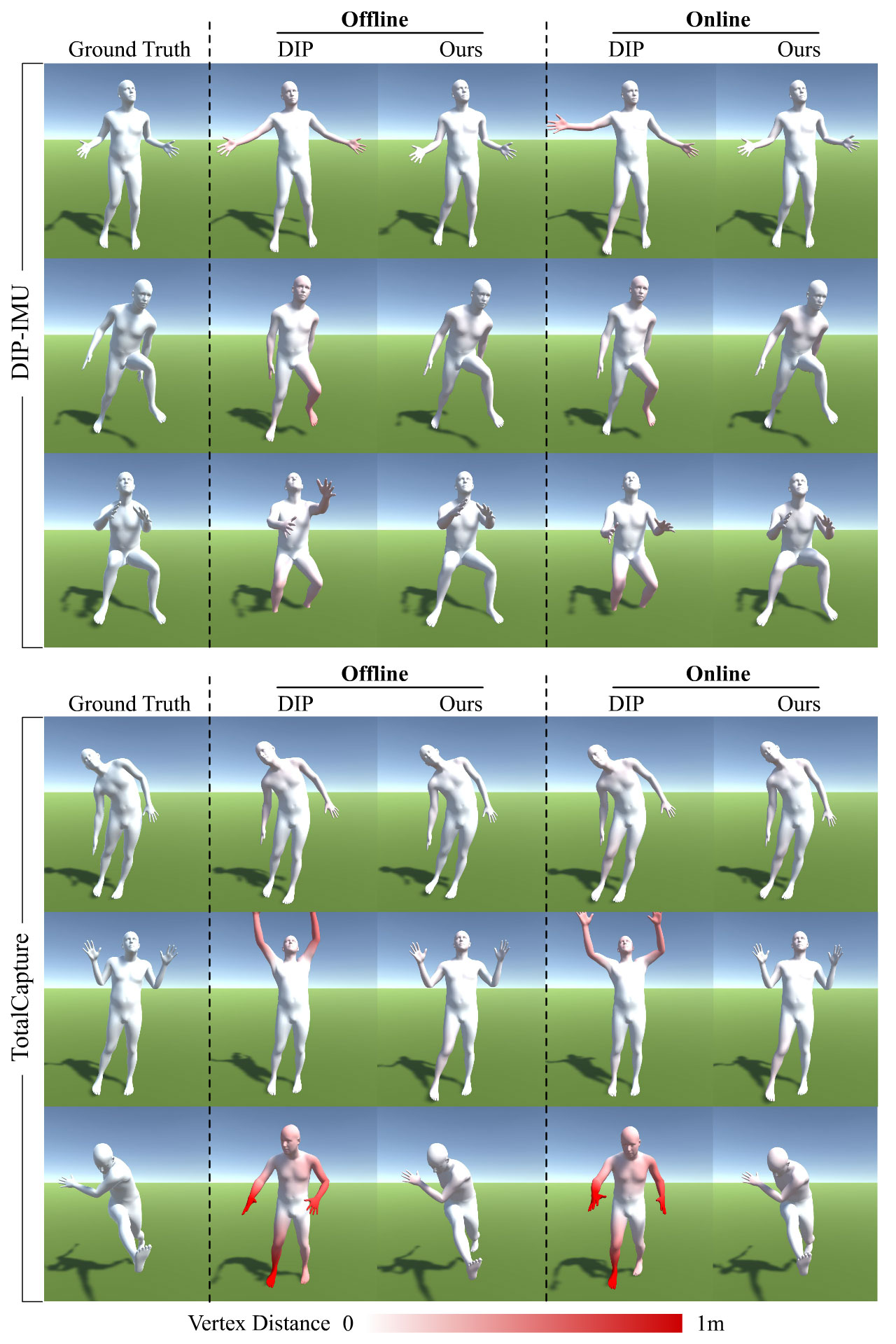}
	\caption{
	    Qualitative comparisons between our method and DIP. 
	    We perform both offline and online comparisons on DIP-IMU (top-three rows) and TotalCapture (bottom-three rows) dataset, and pick some results here. 
	    More comparisons can be found in our video. 
		Each vertex is colored by its distance to the ground truth location. 
		A complete-red vertex means a distance greater than $1\rm{m}$.
	}
	\Description{Qualitative comparison results (in posed body meshes).}
	\label{fig:qual_eval}
\end{figure}
Here, we visually compare offline and online results for pose estimation between our method and the state-of-the-art work DIP.
Some selected frames from the two datasets are shown in Figure \ref{fig:qual_eval} and more comparisons are shown in our video.
From the top three rows in Figure \ref{fig:qual_eval} picked from DIP-IMU dataset, we see that our approach reconstructs the upper and lower body well, while DIP does not accurately estimate the poses of the upper arms and legs.
Since inner joint rotations are not directly measured by any sensor, high ambiguities exist in the mapping from leaf joint rotations to poses of the whole body.
The explicit estimation of joint positions in our pipeline helps the network to make better use of the acceleration information, which is the key to solve such ambiguities.
Thus, our method shows greater accuracy in inner joint rotations.
The following three rows are some challenging poses selected from the TotalCapture dataset.
Our method performs slightly better for the side-bending pose as shown in the top row. 
In the middle row, our method correctly estimates the rotations of upper arms where DIP fails, while both have similar forearm rotations.
We show a challenging case in the last row where DIP fails but we still give a satisfying estimation.
Again, we attribute such superiority of our method to the intermediate joint position estimation tasks which have a regularization effect on the final result.
%
\subsection{Evaluations} \label{sec:evaluations}
%
\subsubsection{Multi-stage pose estimation}
%
\begin{table}[]
	\caption{
	    Evaluation of the multi-stage design for pose estimation using DIP-IMU and TotalCapture dataset.
	    "I" stands for IMU measurements; 
	    "LJ" means leaf joint positions; 
	    "AJ" denotes full joint positions; 
	    and "P" indicates body poses. 
	    We demonstrate the superiority (judging from the SIP error and jitter error) of our multi-stage method which estimates leaf and full joint positions as intermediate tasks.
	}
	\label{tab:ablation_stages}
	\resizebox{\linewidth}{!}{
	\begin{tabular}{ccccc}
		\toprule
		 & \multicolumn{2}{c}{DIP-IMU}  & \multicolumn{2}{c}{TotalCapture} \\
		 \cmidrule(lr){2-3}\cmidrule(lr){4-5}
         & SIP Err ($\rm{deg}$) & Jitter ($10^2\rm{m}/\rm{s}^3$) & SIP Err ($\rm{deg}$)  & Jitter ($10^2\rm{m}/\rm{s}^3$) \\
		\midrule
		I$\to$P    & 14.43 ($\pm$7.77)          &  2.50 ($\pm$3.42)         & 23.16($\pm$9.00)          & 3.34 ($\pm$5.72)          \\
		I$\to$LJ$\to$P  & 14.35 ($\pm$7.75)          &  2.22 ($\pm$3.32)         & 17.71 ($\pm$7.89)          & 2.90 ($\pm$5.09)          \\
		I$\to$AJ$\to$P & 14.29 ($\pm$7.30)          &  1.23 ($\pm$1.82)         & 19.76 ($\pm$8.05)          & 1.60 ($\pm$2.94)          \\
		I$\to$LJ$\to$AJ$\to$P   & \textbf{13.97 ($\pm$6.77)} & \textbf{1.19 ($\pm$1.76)} & \textbf{14.95 ($\pm$6.90)} & \textbf{1.57 ($\pm$2.93)} \\
		\bottomrule
	\end{tabular}}
\end{table}
We demonstrate the superiority of our three-stage pose estimation structure here.
We evaluate the following three variants of our pipeline:
\textit{1)} directly regressing joint rotations from inertial inputs, without any intermediate task;
\textit{2)} estimating joint rotations with the help of intermediate \textit{leaf} joint coordinates regressed from inertial measurements, i.e. combining Pose-S2 and Pose-S3;
\textit{3)} estimating joint rotations with the help of intermediate \textit{all} joint positions, which are directly regressed from inertial measurements, i.e. combining Pose-S1 and Pose-S2.
We compare these variations with our original three-stage method on DIP-IMU and TotalCapture dataset in the offline setting.
As shown in Table \ref{tab:ablation_stages}, estimating joint positions greatly helps with the pose estimation task, which leads to better accuracy and a significant reduction of jitters.
We attribute this to the nonlinearity of human poses in the joint rotation representation, which makes it hard to learn how to extract useful information from the linear accelerations of body joints, resulting in an automatic discard of acceleration data as reported in DIP \cite{DIP}.
Hence, by estimating joint positions as an intermediate task, we make better use of acceleration information due to its linear correlation with positions, and as a result, we do not need any special loss for accelerations where DIP does.
Another observation from Table \ref{tab:ablation_stages} is that the leaf joint positions are helpful for the estimation of other joints. 
We think this is because estimating leaf joint positions is easier than inner ones as their inertia is directly measured by IMUs.
Then, with the help of the leaf joints, it is easier to estimate the inner ones, as the leaf joints are affected by inner ones.
So breaking down such a complex task into easier ones helps to achieve better results.
%
\subsubsection{Fusion-based global translation estimation}
%
\begin{figure}
	\includegraphics[width=\linewidth]{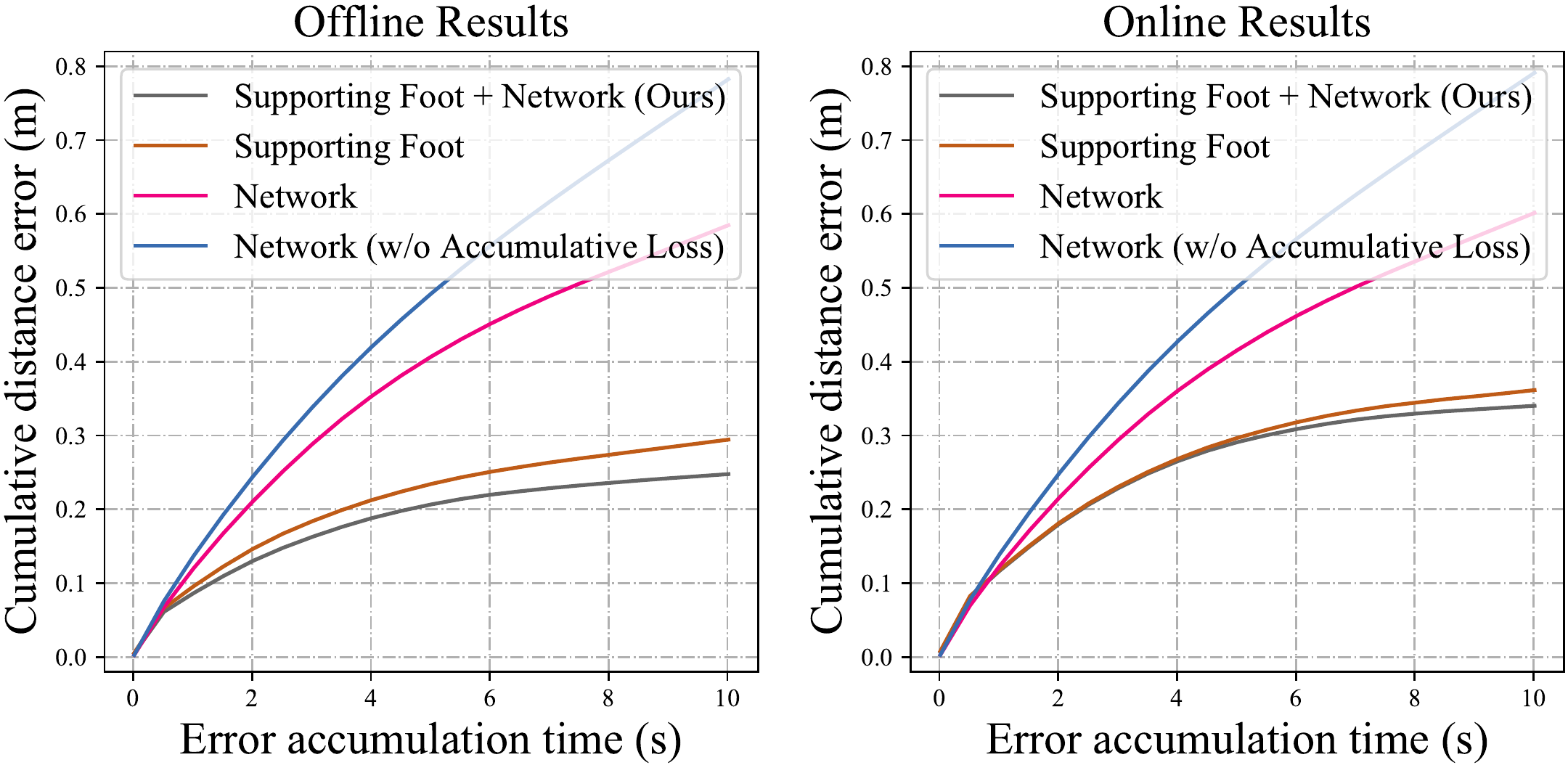}
	\caption{
		Evaluation of global translation estimation. 
		Here we evaluate cumulative distance errors with respect to time. 
		The error of each solution is plotted as a curve. 
		A lower curve means better translation accuracy. 
		We demonstrate the superiority of our fusion-based method and also the effectiveness of the accumulative loss terms (Equation \ref{equ:accloss}) used in the training of \textit{Trans-B2}.
	}
	\Description{Cumulative distance error curves.}
	\label{fig:tran_eval}
\end{figure}
We conduct an evaluation on the global translation estimation task to demonstrate the advantages of our hybrid supporting-foot- and network-based approach.
Each alternative method is evaluated and compared with our fusion-based implementation.
Specifically, we evaluate:
\textit{1)} the supporting-foot-based method in Trans-B1, which means that we only estimate foot-ground contact states and compute the global translations from poses;
\textit{2)} using the neural network to regress translations in Trans-B2, which means that we only estimate global translations from an RNN;
\textit{3)} running both branches and fusing them according to the estimated foot-ground contact probabilities, i.e. our hybrid method.
Additionally, we evaluate \textit{4)} running only the RNN in Trans-B2, but trained with a simple L2 loss that measures the mean square error of the predicted velocities, i.e. using only $\mathcal{L}_{\mathrm{vel}}(1)$ rather than the accumulative loss (Equation \ref{equ:accloss}).
%
%
All the methods above are evaluated on TotalCapture dataset in both offline and online settings.
We draw the cumulative error curve with respect to time in Figure \ref{fig:tran_eval}, i.e. the mean Euclidean distance between the estimated and ground truth global positions with predefined error accumulation time ranging from 0 to 10 seconds.
This mean distance is calculated by first aligning the root of the estimated and ground truth body in one frame, then recording the distance errors in the following $t$ (ranging from 0 to 10) seconds, and finally averaging the errors over all frames in the test dataset.
We should note that there are five subjects in TotalCapture dataset and their leg lengths are acquired by averaging the ground truth lengths of all frames.
As shown in Figure \ref{fig:tran_eval}, our hybrid approach outperforms all the other methods in both offline and online tests.
It indicates that our hybrid method, which computes global motions majorly using the supporting foot and deals with the remaining unsolved movements using a network trained on complementary data, makes up the flaws of both branches and thus shows better results.
However, this benefit is not significant comparing with the method in Trans-B1 because the supporting foot can be detected in most of the test datasets.
So the result is dominated by Trans-B1.
In addition, the RNN trained without the accumulative loss performs worse than the one with such a loss term.
It demonstrates the effectiveness of the accumulative terms which benefit the estimation of long-range movements.
Finally, the errors caused by jittering will be canceled out somehow in the long run, resulting in the flattening trend of the curves.
%
\subsubsection{The network for velocity estimation}
%
\begin{table}[]
	\caption{
	    Evaluation of the network of Trans-B2 on TotalCapture dataset. 
	    RNN and biRNN are compared using the cumulative distance errors with $1\mathrm{s}$ (left) and $5\mathrm{s}$ (right) accumulation time.
	    We demonstrate that RNN is better suited for the velocity estimation task.
	}
	\label{tab:vel_net}
	\resizebox{\linewidth}{!}{
	\begin{tabular}{ccccc}
		\toprule
		\multirow{2}*{Subject} & \multicolumn{2}{c}{Dist Err@$1\mathrm{s}$ ($\mathrm{cm}$)} & \multicolumn{2}{c}{Dist Err@$5\mathrm{s}$ ($\mathrm{cm}$)}   \\
		\cmidrule(lr){2-3}\cmidrule(lr){4-5}
		& biRNN       & RNN             & biRNN       & RNN             \\
		\midrule
		s1    & 14.20 ($\pm$7.51)      & \textbf{11.34} ($\pm$\textbf{5.49})  & 47.74 ($\pm$20.29)     & \textbf{40.07} ($\pm$\textbf{16.73}) \\
		s2    & 12.75 ($\pm$7.41)      & \textbf{9.38} ($\pm$\textbf{7.09})   & 41.64 ($\pm$17.41)     & \textbf{30.04} ($\pm$\textbf{17.23}) \\
		s3    & 15.76 ($\pm$8.38)      & \textbf{12.15} ($\pm$\textbf{6.79})  & 52.45 ($\pm$21.05)     & \textbf{41.32} ($\pm$\textbf{18.12}) \\
		s4    & 22.18 ($\pm$17.13)     & \textbf{20.09} ($\pm$\textbf{14.54}) & 72.18 ($\pm$42.15)     & \textbf{68.40} ($\pm$\textbf{34.31}) \\
		s5    & 18.54 ($\pm$10.53)     & \textbf{13.31} ($\pm$\textbf{6.59})  & 55.42 ($\pm$27.53)     & \textbf{46.93} ($\pm$\textbf{17.81}) \\
		total & 15.50 ($\pm$9.05)      & \textbf{12.18} ($\pm$\textbf{7.36})  & 50.77 ($\pm$22.80)     & \textbf{41.49} ($\pm$\textbf{19.28}) \\
		\bottomrule
	\end{tabular}}
\end{table}
We conduct an evaluation of the network structure used for global translation estimation.
We compare \textit{1)} a \textit{unidirectional} RNN and \textit{2)} a biRNN using 20 past, 1 current, and 5 future frames on TotalCapture dataset in the online setting.
As shown in Table \ref{tab:vel_net}, RNN outperforms biRNN on distance errors for both 1 and 5 seconds of accumulation time.
Due to the long-term historical dependency of human movements, RNN can utilize all historical information while biRNN can only acquire a limited number of past frames when running in an online manner.
On the other hand, RNN only takes one forward pass for a new frame, while biRNN needs to process all the 26 frames in the window forward and backward, which is about 50 times slower.
To increase the estimation accuracy, more frames are needed for biRNN, which further reduce the runtime performance.
Hence, RNN is a more suitable choice for the global translation estimation.
%
\subsubsection{Cross-layer connections of IMU measurements}
%
\begin{table}[]
	\caption{
	    Evaluation of the cross-layer connections of IMU measurements using DIP-IMU and TotalCapture dataset.
	    We compare our original pose estimation pipeline with removing the inertial inputs in Pose-S2 and Pose-S3.
	    We demonstrate the superiority of the cross-layer connections of IMU measurements which help with full joint position estimation and IK solving.
	}
	\label{tab:ablation_imus}
	\resizebox{\linewidth}{!}{
	\begin{tabular}{ccccc}
		\toprule
		 & \multicolumn{2}{c}{DIP-IMU}  & \multicolumn{2}{c}{TotalCapture} \\
		 \cmidrule(lr){2-3}\cmidrule(lr){4-5}
         & SIP Err ($\rm{deg}$) & Mesh Err ($\rm{cm}$) & SIP Err ($\rm{deg}$)  & Mesh Err ($\rm{cm}$) \\
		\midrule
		S2 w/o IMUs   & 17.06 ($\pm$7.29)          &  6.44 ($\pm$3.38)         & 18.51 ($\pm$7.31)          & 6.84 ($\pm$3.61)          \\
		S3 w/o IMUs   & 15.66 ($\pm$7.53)          &  6.50 ($\pm$3.51)         & 15.75 ($\pm$7.18)          & 6.83 ($\pm$3.67)          \\
		Ours   & \textbf{13.97 ($\pm$6.77)} & \textbf{5.83 ($\pm$3.21)} & \textbf{14.95 ($\pm$6.90)} & \textbf{6.36 ($\pm$3.47)} \\
		\bottomrule
	\end{tabular}}
\end{table}
We demonstrate the effectiveness of the cross-layer connections of IMU measurements in the pose estimation task, i.e. the impact of the inertial inputs in Pose-S2 and Pose-S3.
Specifically, we evaluate two variations of our pose estimation pipeline:
\textit{1)} removing the IMU measurements in the input of Pose-S2, i.e. estimating full joint positions only from the leaf joint positions;
\textit{2)} removing the IMU measurements in the input of Pose-S3, i.e. solving the IK problem only from full joint positions.
As shown in Table \ref{tab:ablation_imus}, knowing leaf joint inertia helps with the estimation of inner joint positions and rotations.
We attribute this to the kinematic structure of human bodies, where leaf joints are affected by the movements of inner joints.
Thus, by leveraging the leaf joint inertia via skip connections, Pose-S2 and Pose-S3 can achieve better accuracy.
\subsection{Live Demo} \label{sec:livedemo}
%
We implement a real-time live mocap system and show its results in the accompanying video.
Concretely, we attach six IMUs onto the corresponding body parts and read the raw inertial measurements via wireless transmission.
We calibrate the received data (see Appendix \ref{app:imu}) and predict the poses and global translations, which are finally used to drive a 3D human model in real-time.
From the live demos, we can see that our system handles strong occlusion, wide range motion space, dark and outdoor environment, close interaction, and multiple users with high accuracy and real-time performance.
%
\subsection{Limitations}
%
\subsubsection{On the hardware side}
%
In the inertial sensor, a magnetometer is used to measure directions. 
However, the magnetometer is easily affected by the magnetic field of the environment. 
As a result, it cannot work in an environment with spatially or temporally varying magnetic fields. 
%
%
Our current sensor uses wireless transmission to deliver signals to our laptop, which does not allow the sensor to be far from the laptop (10 meters at most in our experiments). 
Hence, to record wide range motions, we still need to carry the laptop and follow the performer. 
The transmission problem should also be solved to extend the usage of the techniques based on inertial sensors.
%
\subsubsection{On the software side}
%
As a data-driven method, our approach also suffers from the generalization problem. 
It cannot handle motions that are largely different from the training dataset, e.g., splits and other complex motions that can only be performed by professional dancers and players.
Our supporting-foot-based method relies on the assumption that the supporting foot has no motion in the world coordinate space, which is not true for motions like skating and sliding. 

\section{Conclusion}
%
This paper proposes TransPose, a 90-fps motion capture technique based on 6 inertial sensors, which reconstructs full human motions including both body poses and global translations. 
In this technique, a novel multi-stage method is proposed to estimate body poses based on the idea of reformulating the problem with two intermediate tasks of leaf and full joint position estimations, which leads to a large improvement over the state-of-the-art on accuracy, temporal consistency, and runtime performance.
For global translation estimation, a fusion technique is used to leverage a purely data-driven method and a method involving the motion rules of the supporting foot and a kinematic body model. 
By combining data-driven and motion-rule-driven, the challenging problem of global translation estimation from noisy sparse inertial sensors is solved in real-time for the first time to the best of our knowledge.
Extensive experiments with strong occlusion, wide range motion space, dark and outdoor environment, close interaction, and multiple users demonstrate the robustness and accuracy of our technique.
\begin{acks}
We would like to thank Yinghao Huang and the other DIP authors for providing the SMPL parameters for TotalCapture dataset and the SIP/SOP results. 
We would also like to thank Associate Professor Yebin Liu for the support on the IMU sensors.
We also appreciate Hao Zhang, Dong Yang, Wenbin Lin, and Rui Qin for the extensive help with the live demos.
We thank Chengwei Zheng for the proofreading, and the reviewers for their valuable comments.
This work was supported by the National Key R\&D Program of China 2018YFA0704000, the NSFC (No.61822111, 61727808) and Beijing Natural Science Foundation (JQ19015). Feng Xu is the corresponding author.
\end{acks}

\bibliographystyle{ACM-Reference-Format}
\bibliography{bibtex}

\appendix
\section{Sensor Preprocessing} \label{app:imu} 
%
\begin{figure*}[ht]
	\includegraphics[width=\linewidth]{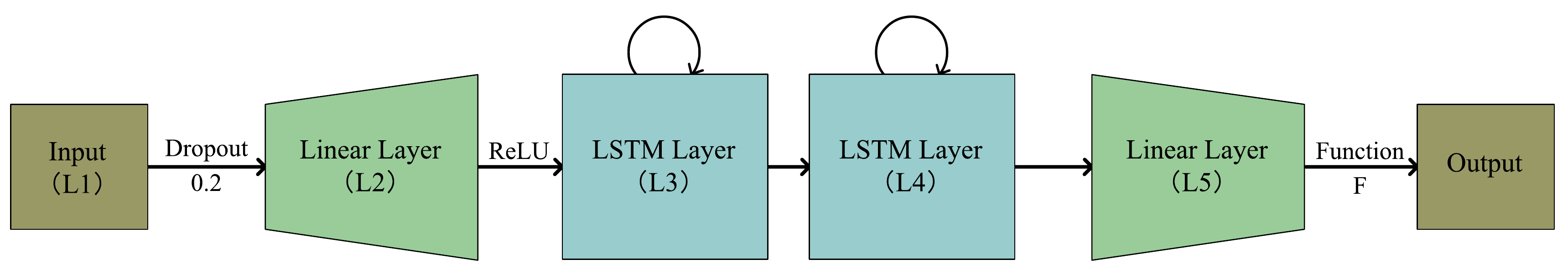}
	\caption{
	    Detailed structures for each network of our pipeline. 
	    "L1" $\cdots$ "L5" represent the output widths of the layers. 
	    "F" represents the additional function applied after the last linear layer. 
	    Values of these parameters for each network are shown in Table \ref{tab:net}.
    }
    \Description{Structures of each network.}
	\label{fig:net}
\end{figure*}
Since each inertial sensor has its own coordinate system, we need to \textit{1)} firstly transform the raw inertial measurements into the same reference frame, which is referred to as \textit{calibration}, 
and then \textit{2)} transform the leaf joint inertia into the root's space and rescale it to a suitable size for the network input, which is referred to as \textit{normalization}.
The sensors can be placed with arbitrary rotations during setup, and our method automatically computes the transition matrices for each sensor before capturing the motion.
This process requires the subject to keep in T-pose for a few seconds.
In this section, we explain the details of the sensor preprocessing in our method, including the calibration (Section \ref{app:calibration}) and the normalization (Section \ref{app:normalization}).
%
\subsection{Calibration} \label{app:calibration}
%
An inertial measurement unit (IMU) outputs acceleration data relative to the sensor coordinate frame $F^S$ and orientation data relative to the global inertial coordinate frame $F^I$.
We define the coordinate frame of the SMPL \cite{SMPL} body model as $F^M$, and the basis matrix of $F^S, F^I, F^M$ as $\boldsymbol{B}^S, \boldsymbol{B}^I, \boldsymbol{B}^M$ respectively, where each consists of three column basis vectors.
Before capturing motions, we firstly put an IMU with the axes of its sensor coordinate frame $F^S$ aligned with the corresponding axes of the SMPL coordinate frame $F^M$, i.e. to place the IMU with its $x$-axis left, $y$-axis up and $z$-axis forward in the real world. 
Then, the orientation measurement $\boldsymbol P^{IM}$ can be regarded as the transition matrix from $F^I$ to $F^M$:
\begin{equation} \label{equ:app1}
	\boldsymbol{B}^M = \boldsymbol{B}^I\boldsymbol P^{IM} \mathrm{.}
\end{equation}
Next, we put each IMU onto the corresponding body part with arbitrary orientations and keep still in a predefined pose (such as the T-pose) with known leaf-joint and pelvis orientations $\boldsymbol R_M^{\mathrm{bone}[i]} (i=0,1,\cdots,5)$  (relative to $F^M$) for several seconds. 
We read the IMU measurements and calculate the average acceleration (relative to $F^S$) and orientation (relative to $F^I$) of each sensor as $\boldsymbol a_S^{\mathrm{sensor}[i]}$  and $\boldsymbol R_I^{\mathrm{sensor}[i]} $  respectively. 
We represent the rotation offsets between the sensors and the corresponding bones as $\boldsymbol R_I^{\mathrm{offset}[i]}$ due to the arbitrarily oriented placement of the sensors and the assumption that the angles between muscles and bones are constants.
We then have:
\begin{equation} \label{equ:app2}
	\boldsymbol R_I^{\mathrm{bone}[i]} = \boldsymbol R_I^{\mathrm{sensor}[i]} \boldsymbol R_I^{\mathrm{offset}[i]} \mathrm{,}
\end{equation}
where $\boldsymbol R_I^{\mathrm{bone}[i]}$ is the absolute orientation of bone $i$ in the coordinate frame $F^I$. 
For any given pose, the absolute  orientation of bone $i$ is equivalent in two coordinate frames $F^M$, $F^I$:
\begin{equation}  \label{equ:app3}
	\boldsymbol B^M \boldsymbol R_M^{\mathrm{bone}[i]} = \boldsymbol B^I \boldsymbol R_I^{\mathrm{bone}[i]} \mathrm{.}
\end{equation}
Combining Equation \ref{equ:app1}, \ref{equ:app2}, and \ref{equ:app3}, we can get:
\begin{equation} \label{equ:app4}
	\boldsymbol R_I^{\mathrm{offset}[i]} = (\boldsymbol R_I^{\mathrm{sensor}[i]})^{-1}\boldsymbol P^{IM}\boldsymbol R_M^{\mathrm{bone}[i]} \mathrm{.}
\end{equation}
For accelerations, we first transform the measurements in the sensor local frame $F^S$ to the global inertial frame $F^I$ as:
\begin{equation} \label{equ:app5}
	\boldsymbol a_I^{\mathrm{bone}[i]} = \boldsymbol a_I^{\mathrm{sensor}[i]} = \boldsymbol R_I^{\mathrm{sensor}[i]} \boldsymbol a_S^{\mathrm{sensor}[i]} \mathrm{,}
\end{equation}
where $\boldsymbol a_I^{\mathrm{bone}[i]}$ and $\boldsymbol a_I^{\mathrm{sensor}[i]}$ are the accelerations of bone $i$ and the IMU attached to it in $F^I$, respectively. 
They are equal because we assume the IMUs do not move relatively once mounted.
Due to the sensor error and the inaccuracy of $\boldsymbol P^{IM}$, there is a constant offset in the global acceleration.
Thus, we add an offset $\boldsymbol a_M^{\mathrm{offset}[i]}$ to the coordinate frame $F^M$ as:
\begin{equation} \label{equ:app6}
	\boldsymbol B^M(\boldsymbol a_M^{\mathrm{bone}[i]} + \boldsymbol a_M^{\mathrm{offset}[i]}) = \boldsymbol B^I \boldsymbol a_I^{\mathrm{bone}[i]} \mathrm{.}
\end{equation}
Since the subject keeps still during calibration, $\boldsymbol a_M^{\mathrm{bone}[i]} = \boldsymbol 0$. 
We then combine Equation \ref{equ:app1}, \ref{equ:app5}, and \ref{equ:app6} to get:
\begin{equation} \label{equ:app7}
	\boldsymbol a_M^{\mathrm{offset}[i]} = (\boldsymbol P^{IM})^{-1} \boldsymbol R_I^{\mathrm{sensor}[i]} \boldsymbol a_S^{\mathrm{sensor}[i]} \mathrm{.}
\end{equation}
Leveraging the known $\boldsymbol{P}^{IM}$, $\boldsymbol R_I^{\mathrm{offset}[i]}$, and $\boldsymbol a_M^{\mathrm{offset}[i]}$ estimated in the pre-computation step, we can now perform motion capture.
Specifically, we calculate $\boldsymbol R_M^{\mathrm{bone}[i]}$ and $\boldsymbol a_M^{\mathrm{bone}[i]}$ (denoted as $\boldsymbol{\bar{R}}_i$ and $\boldsymbol{\bar{a}}_i$ in short) per frame, and feed them into our model after the normalization process described in Section \ref{app:normalization}. 
%
\subsection{Normalization} \label{app:normalization}
%
We explain our normalization process in this section.
For each frame, the raw inputs are accelerations
$[\boldsymbol {\tilde a}_{\mathrm{root}}$, $\boldsymbol {\tilde a}_{\mathrm{lleg}}$, $ \boldsymbol {\tilde a}_{\mathrm{rleg}}$, $\boldsymbol {\tilde a}_{\mathrm{head}}$, $\boldsymbol {\tilde a}_{\mathrm{larm}}$, $\boldsymbol {\tilde a}_{\mathrm{rarm}}] \in \mathbb{R}^{3 \times 6}$
and rotations
$[\boldsymbol {\tilde R}_{\mathrm{root}}$, $\boldsymbol {\tilde R}_{\mathrm{lleg}}$, $\boldsymbol {\tilde R}_{\mathrm{rleg}}$, $\boldsymbol {\tilde R}_{\mathrm{head}}$, $\boldsymbol {\tilde R}_{\mathrm{larm}}$, $\boldsymbol {\tilde R}_{\mathrm{rarm}}] \in \mathbb{R}^{3 \times 3 \times 6}$
measured by the IMUs.
We transfer these measurements from their own coordinate frames to the SMPL reference frame, obtaining $\boldsymbol {\bar a}$ and $\boldsymbol {\bar R}$ as described in Section \ref{app:calibration}.
Then, we align leaf joint inertial measurements with respect to the root as:
\begin{equation}
	\boldsymbol a_{\mathrm{leaf}} = \boldsymbol {\bar R}_{\mathrm{root}}^{-1}(\boldsymbol {\bar a}_{\mathrm{leaf}} - \boldsymbol {\bar a}_{\mathrm{root}}) \mathrm{,}
\end{equation}
\begin{equation}
	\boldsymbol R_{\mathrm{leaf}} = \boldsymbol {\bar R}_{\mathrm{root}}^{-1}\boldsymbol {\bar R}_{\mathrm{leaf}} \mathrm{,}
\end{equation}
and normalize root joint measurements as:
\begin{equation}
	\boldsymbol a_{\mathrm{root}} = \boldsymbol {\bar R}_{\mathrm{root}}^{-1}\boldsymbol {\bar a}_{\mathrm{root}} \mathrm{,}
\end{equation}
\begin{equation}
	\boldsymbol R_{\mathrm{root}} = \boldsymbol {\bar R}_{\mathrm{root}} \mathrm{.}
\end{equation}
Finally, we rescale the accelerations to a suitable size for neural networks.

\section{Network Details} \label{app:net}
%
\begin{table}[t]
	\caption{
	    Details of each network in our pipeline. 
	    "L1" $\cdots$ "L5" are the output widths of the corresponding layers as shown in Figure \ref{fig:net}. 
	    "F" is the function applied after the output layer. 
	    "Bidirectional" represents whether the two LSTM layers are bidirectional.
	} 
	\label{tab:net}
	\resizebox{\linewidth}{!}{
	\begin{tabular}{cccccccc}
		\toprule
		& L1 & L2 & L3 & L4 & L5 & F & Bidirectional  \\
		\midrule
	 	Pose-S1 & 72 & 256 & 256 & 256 & 15 & - & true \\
		Pose-S2 & 87 & 64 & 64 & 64 & 69 & - & true \\
		Pose-S3 & 141 & 128 & 128 & 128 & 90 & - & true \\
		Trans-B1 & 87 & 64 & 64 & 64 & 2 & sigmoid & true \\
		Trans-B2 & 141 & 256 & 256 & 256 & 3 & - & false \\
		\bottomrule
	\end{tabular}}
\end{table}
We introduce the detailed structures for all the networks in our pipeline here. 
As shown in Figure \ref{fig:net}, each network contains a direct 20\% dropout on the input, followed by a linear and a ReLU operation that map the input to the dimension of the LSTM layers. 
Then two LSTM layers with the same width process the data and a linear operation maps it to the output dimension. 
Particularly, we apply a sigmoid function on the output in the network of Trans-B1 that estimates foot-ground contact probabilities.
The output widths of the layers for each network are listed in Table \ref{tab:net}.

\section{Acceleration Synthesis} \label{app:syn}
%
\begin{figure} 
	\includegraphics[width=\linewidth]{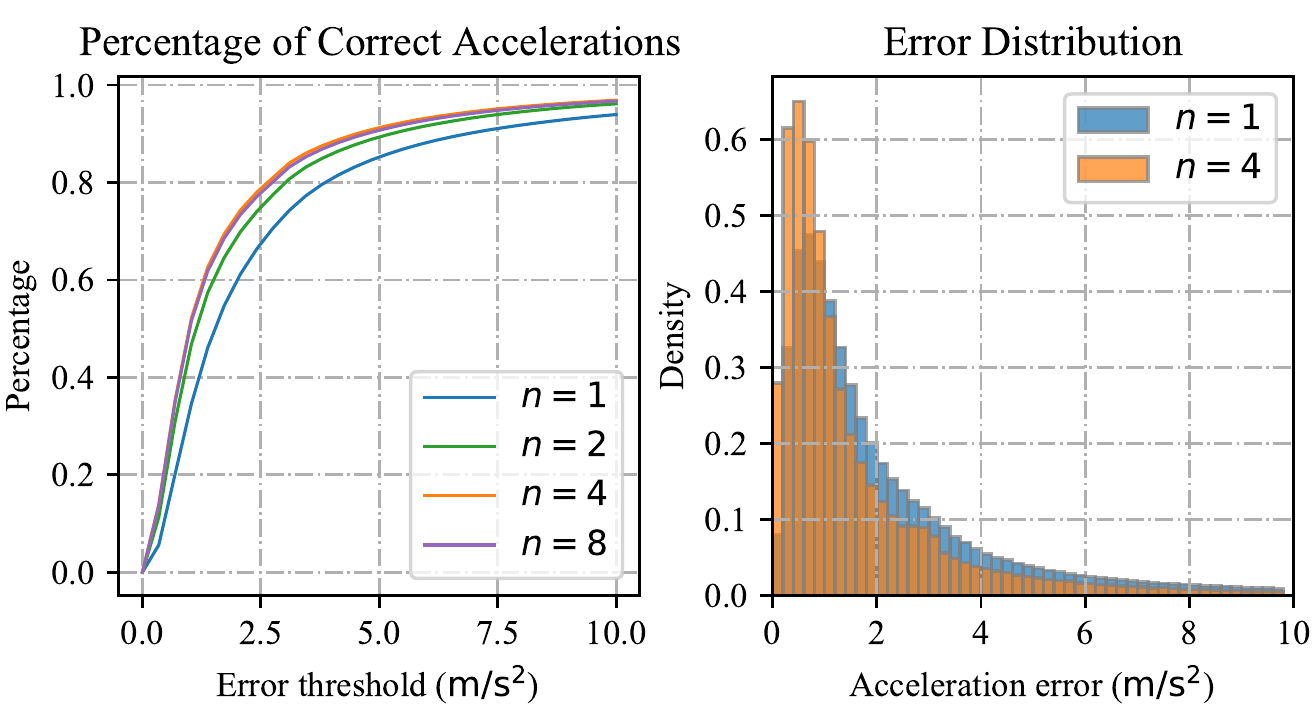}
	\caption{
	    Influences of the smoothing factor $n$ in Equation \ref{equ:syn}. 
	    We evaluate the \textit{percentage of correct synthetic accelerations} on TotalCapture dataset using different smoothing factors $n$ and demonstrate that a value of $n=4$ produces accelerations with the closest distribution to the measurements of real sensors.
	}
	\Description{Percentage of correct acceleration curves and an error distribution histogram.}
	\label{fig:pck}
\end{figure}
We need to synthesize leaf joint inertia for AMASS dataset as stated in Section \ref{sec:trainingdata}.
Here we show the influence of the smoothing factor $n$ in Equation \ref{equ:syn} and demonstrate our advantages to use $n=4$ over other works \cite{SIP, DIP, Malleson2019, Malleson2017} which assume $n=1$. 
In order to demonstrate the purpose of the smoothing factor, we re-synthesize accelerations for TotalCapture dataset using the ground truth poses and translations, and compare the synthetic accelerations with real IMU measurements. 
We evaluate the {\it percentage of correct accelerations} under some selected error thresholds and plot the curves and the error distributions in Figure \ref{fig:pck}. 
The results show that a smoothing factor of $n=4$ produces accelerations most similar to real sensor measurements, while the classical finite difference method where $n = 1$ produces less accurate results due to the jitters of leaf joints.
Therefore, we take $n=4$ in our work.

\end{document}